\documentclass[aps,superscriptaddress,preprint,showpacs,nofootinbib,floatfix]{revtex4}
\usepackage{bm}
\usepackage{dcolumn}
\usepackage{epsfig}
\usepackage{pstricks}
\begin{document}
\title{Pion re-scattering operator in the S-matrix approach}
\author{V.~Malafaia}
\affiliation{Instituto Superior T\'ecnico, Centro de F\'{\i}sica
Te\'orica de Part\'\i culas and Department of Physics, Av. Rovisco
Pais, 1049-001 Lisboa, Portugal }
\author{J.~Adam, Jr.}
\affiliation{Nuclear Physics Institute, \v{R}e\v{z} near Prague,
    CZ-25068, Czech Republic}
\author{M.T.~Pe\~na}
\affiliation{Instituto Superior T\'ecnico, Centro de F\'{\i}sica
Te\'orica de Part\'\i culas and Department of Physics, Av. Rovisco
Pais, 1049-001 Lisboa, Portugal }

\date{\today}
\begin{abstract}
The pion re-scattering operator for pion production, derived recently in
time-ordered perturbation theory, is compared with the one following from the
simple S-matrix construction. We show that this construction is equivalent to
the on-shell approximation introduced in previous papers. For a realistic
$NN$ interaction, the S-matrix approach, and its simplified fixed
threshold-kinematics version, work well near threshold.
\end{abstract}
\pacs{13.60.Le, 25.40.Ve, 21.45.+v, 25.10.+s}


\maketitle

\section{Introduction}

The detailed analysis of the irreducible pion re-scattering operator was
recently performed \cite{MP} for the reaction $pp \rightarrow pp \pi^0$. The
pion re-scattering is certainly part of the pion production mechanism, but
its importance relative to other contributions varies considerably dependent
on the approximations made in evaluation of the effective operators (see
Ref.~\cite{MP} and references therein). The nature and extent  of this
uncertainty are re-examined in this paper.

To this end we deal with retardation effects in the exchanged pion
propagator, i.e., its energy dependence, as well as with the energy
dependence of the $\pi$N scattering amplitude in the vertex, from which the
produced pion is emitted. While the approximations employed in a previous
paper \cite{MP} yield rather different results, we show here that the
deviations between them are significantly reduced when the approximations are
applied consistently in the whole effective operator.

We also show that the S-matrix approach, which has been successfully used
below pion production threshold,  yields also above threshold results rather
close to those obtained with the energy-dependent operator following from
time-ordered perturbation theory.

The paper is organized as follows: after this Introduction, section II
describes the S-matrix technique for deriving effective nuclear
quantum-mechanical operators, section III describes the results and section
IV presents a summary and conclusions.

\section{Pion re-scattering operator}

To derive the effective pion production operators, and other effective
nuclear operators in general, one starts from the relativistic (effective)
Lagrangian written in terms of hadronic fields. The interactions mediated by
meson exchanges before and after the production reaction takes place are
included in the effective $NN$ (and nucleon-meson) interaction, while from
the irreducible parts connected to the reaction mechanism (e.g., pion
production) one obtains effective operators, whose expectation values are to
be evaluated between the initial and final nucleonic wave functions. One
aims is to get such effective operators consistent with the realistic
description of the $NN$ interaction, which can then be used in studies of the
corresponding reactions not only on the simplest (one or two-nucleon)
systems, but preferably also on heavier nuclei.

The covariant techniques based on the Bethe-Salpeter equation or its
quasipotential re-arrangements are these days practically manageable only
below meson production threshold. However, above the threshold the dressings
of the single hadron propagators and interaction vertices via the meson loops
have to be included explicitly. For this reason the construction of the
production operator is so far realized mostly in the Hamiltonian
quantum-mechanical framework (usually non-relativistic, or with leading
relativistic effects included perturbatively within the decomposition in
powers of $p/m$, where $p$ is typical hadronic momentum and $m$ is the
nucleon mass).

The derivation of the nuclear effective operators below the meson production
threshold within the Hamiltonian framework -- leading to hermitian and energy
independent $NN$ and $3N$ potentials and conserved e.m.\ and partially
conserved weak current operators -- can be done in many different ways (see
discussion in Ref.~\cite{ATA} and references therein). At the
non-relativistic order the results are determined uniquely. As for the
leading order relativistic contributions, they were shown to be unitarily
equivalent. The unitary freedom allows to choose the $NN$ potentials to be
static (in the c.m.\ frame of two-nucleon system) and identify them with the
successful static semi-phenomenological potentials.

Also above the threshold the static limit is commonly employed, since more
elaborate descriptions which include the mesonic retardation and loop effects
are technically considerably more complex \cite{elster,schwamb}, especially
for systems of more than two nucleons. Both the static approaches and the
ones including ``retardation''
typically consider contributions of several one-meson exchanges
and the potentials are fitted to describe the data. It is therefore difficult
to assess how well do they approximate the covariant amplitudes
(corresponding to the same values of physical masses and coupling constants)
which are so far outside the scope of existing calculations schemes, but
which we believe do provide in principle the consistent description of the
considered reactions.

Thus, the ultimately exact approach to the description of the pion production
(and in particular of the pion re-scattering contribution) would be either
the covariant Bethe-Salpeter or quasipotential frameworks (extended above the
pion threshold) or the quantum mechanical coupled-channel technique including
retardation. In these approaches one has to treat the non-hermitian
energy-dependent $NN$ interaction (fitted to the data also above pion
production threshold) and consider the effects of renormalization of
vertices, masses and wave functions via meson loops.

In this paper we rather (following Refs.~\cite{H}) numerically estimate the
range of the predictions from several commonly used simplifying
approximations, and compare them to the result obtained from the reduction of
the corresponding covariant Feynman diagrams for the pion re-scattering
operator. This reduction coincides with the time-ordered perturbation
theory\cite{MP}.

\subsection{Factorization of the effective re-scattering operator}

\begin{figure}
\includegraphics[width=.58\textwidth,keepaspectratio]{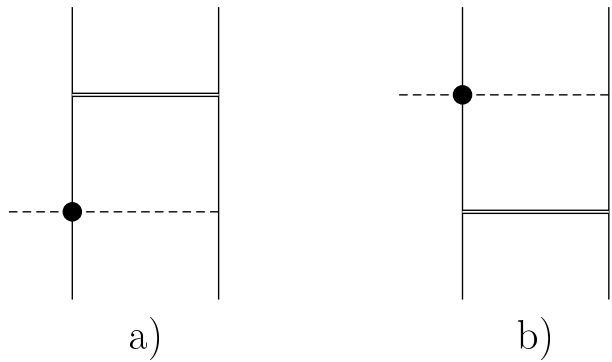}
\caption{Feynman diagrams for pion re-scattering. The pion  field is
represented by a dashed line, the $NN$ interaction by solid double line and
the nucleons by solid lines.}
\label{diagrams1}
\end{figure}

In a previous paper \cite{MP} we made the connection to the usual
representation of the pion re-scattering operator for non-relativistic
calculations by following the approach of Refs.~\cite{H}. We started from the
covariant two-nucleon Feynman amplitudes including final and initial state
interaction (FSI and ISI, respectively), shown in Figs.~\ref{diagrams1}a and
\ref{diagrams1}b.

\begin{figure}
\includegraphics[width=.58\textwidth,keepaspectratio]{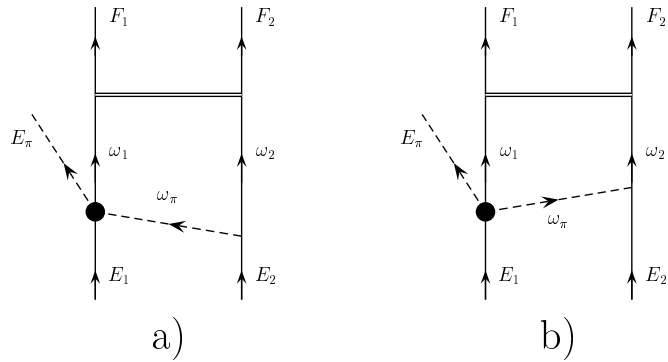}
\caption{The two time-ordered diagrams for FSI considered here,
additional stretched box diagrams are neglected.}
\label{diagrams2}
\end{figure}

To obtain the effective re-scattering operator the negative energy
contributions in the nucleon propagators (to be included in the complete
calculations) were neglected. By integrating subsequently over the energy of
the exchanged pion the resulting Feynman amplitudes were transformed into
those following from the time-ordered perturbation theory. We have shown in
Ref.~\cite{MP} that the irreducible ``stretched box diagrams'' (i.e., those
with more than one meson in flight in the intermediate states) give a very
small contribution and can be therefore also neglected. Thus, the full
covariant amplitude is in the lowest order Born approximation well
approximated  by the product of the $NN$ potential and the effective pion
re-scattering operator, which can be extracted from these time-ordered
diagrams (Fig.~\ref{diagrams2}).

The effective pion re-scattering operator  was in Ref.\cite{MP} factorized
into an effective pion re-scattering  vertex $\tilde{f}$ and an effective
pion propagator $G_\pi$.  For the diagram with FSI (Fig.~\ref{diagrams2})
this factorization reads
\begin{eqnarray}
{\cal M}_{FSI}&=& \int \frac{d^3 q'}{(2\pi)^3}\,
 V_\sigma\, \frac{1}{F_1+F_2- \omega_1- \omega_2 + i\varepsilon}\,
 \hat{O}_{rs}  \, , \label{afsi}\\
  \hat{O}_{rs}  &=& -\frac{1}{2\omega_\pi}
 \left[ \frac{f(\omega_\pi)}{E_2-\omega_2-\omega_\pi}+
 \frac{f(-\omega_\pi)}{E_1-\omega_1-E_\pi-\omega_\pi} \right]
= \frac{1}{2}\, \tilde{f}\, G_\pi \, , \label{mfsi}\\
  \tilde{f} &=& \frac{1}{\omega_\pi}\left[
  (E_1- \omega_1- E_\pi- \omega_\pi) f(\omega_\pi)+
  (E_2- \omega_2- \omega_\pi) f(-\omega_\pi)\right] \, , \label{vtil}\\
  G_\pi &=& - \frac{1}{(E_1- \omega_1- E_\pi- \omega_\pi)(E_2- \omega_2-
  \omega_\pi)} \, ,\label{gpi}\\
  V_\sigma&=& \frac{1}{2\omega_\sigma} \left[
\frac{1}{F_2-\omega_2-\omega_\sigma}+
\frac{1}{F_1-\omega_1-\omega_\sigma} \right] \, .
\label{vsig}
\end{eqnarray}
where, adopting the notation of Ref.~\cite{MP}, $\vec{q}^{\,
\prime}$ is the momentum of the exchanged pion, $\omega_\pi^2=
m_\pi^2+ \vec{q}^{\, \prime 2}$ is its on-mass-shell energy,
$f(\omega_\pi)$ is the product of the $\pi N$ amplitude with the
$\pi NN$ vertex  (as in Ref.~\cite{MP} the standard $\chi$PT
re-scattering vertex is employed here), $E_i, \omega_i, F_i$ are
the on-mass-shell energies of the i-th nucleon in the initial,
intermediate and final state, respectively, $E_\pi$ is the energy
of the produced pion, $E_1+E_2= F_1+ F_2+ E_\pi$.

The inclusion of some pieces of the integrand of Eq.~(\ref{afsi})
into the propagator $G_\pi$ and of others in the modified vertex
$\tilde f$ is somewhat arbitrary. The appearance of the unusual
effective propagator $G_\pi$ and the effective vertex $\tilde{f}$
is the result of combining {\em two} time-ordered diagrams with
different energy dependence  into a {\em single} effective
operator.

The $NN$ interaction is in Fig.~\ref{diagrams2} and  in Eq.~(\ref{afsi})
simulated by a simple $\sigma$-exchange potential. Though not realistic,
this interaction suffices for model studies of approximations
employed in derivations of the effective pion re-scattering operator, as done
in references \cite{H}.
Since some results do depend on the behavior of the $NN$ scattering wave function,
in particular in the region of higher relative momenta, we perform
our calculations (as in Ref.~\cite{MP}) also with $V_\sigma$ replaced by
a full $NN$  T-matrix, generated from realistic Bonn B potential.

We note that the meson poles are not neglected in the integration
over the energy $Q'_0$ of the exchanged pion, which generates
Eq.~(\ref{afsi}). A result similar to Eq.~(\ref{afsi}) can also be
obtained for the amplitude with the initial state interaction
(ISI).
The two amplitudes differ however in the contribution from the pion poles
to the remaining integration over the three-momentum.
For the amplitude with FSI there are no such poles. However, for
the ISI case there are values of the exchanged pion three-momentum
for which the propagator $G_\pi$  has poles. These poles have been
considered in all our numerical calculations for the cross
section. As we will see, they are one of the main reasons for
deviations between several approximations and the reference results
calculated from Eqs.~(\ref{afsi}-\ref{vsig}).

It is worth mentioning that although the FSI and ISI diagrams graphically
separate the $NN$ interaction and the pion re-scattering part (when the
stretched boxes are neglected), they do not define a {\em single} effective
operator (as a function of nucleon three-momenta and the energy of emitted
pion). Since in these time-ordered diagrams energy is not conserved at
individual vertices, each of these diagrams defines a different off-energy
shell extension of the pion re-scattering amplitude. This is an unpleasant
feature, since one would have to make an analogous construction for diagrams
with both FSI and ISI. Moreover, one would have to repeat the whole analysis for
systems of more than two nucleons. Only after the on-shell approximation is
made (in the next subsection), the pion re-scattering parts of FSI and ISI
diagrams coincide and one can identify them with a single effective
re-scattering operator.

\subsection{Re-scattering operator in the S-matrix technique}

The S-matrix technique is a simple prescription to derive the effective
nuclear operators from the corresponding covariant Feynman diagrams
\cite{ATA}. For electromagnetic operators and also for $NN$ and $3N$
potentials the S-matrix approach reproduces the results of more laborious
constructions, based on time-ordered or non-relativistic diagram techniques.

The two-nucleon effective operators are by definition identified with the
diagrams describing the irreducible mechanism of the corresponding reaction.
The only exception are the nucleon Born diagrams from which the iteration of
the one-nucleon operator has to be subtracted. The operators of the nuclear
e.m. and weak currents, as well as the pion absorption operators and
nuclear potentials, are obtained by a straightforward non-relativistic
reduction of the corresponding Feynman diagrams, in which the intermediate
particles are off-mass-shell and energy is conserved at each vertex:
therefore the derived effective operators are also defined
on-energy-shell. The nuclear currents and other transition operators are
defined to be consistent with a hermitian energy independent $NN$ potential,
which has the usual one boson exchange form employed in realistic models of
$NN$ interaction, and can be used also in systems of more than two nucleons.
This approach is well defined and understood below the meson production
threshold, but as a simple tool it is  employed also above the threshold, for
instance in Refs.~\cite{PRS,ASP}, namely to derive the Z-diagram operators.

For the $\sigma$ exchange potential the S-matrix technique in the lowest
order of non-relativistic reduction yields
\begin{equation}
V_\sigma \rightarrow V_\sigma^{on}= - \frac{1}{m_\sigma^2+
\vec{q}_\sigma^{\, 2}- \Delta^2} \, ,
\label{vsigos}
\end{equation}
with $\Delta=\Delta_1= -\Delta_2$ and $\Delta_i= \epsilon'_i-
\epsilon_i$; $\epsilon'_i$ and $\epsilon_i$ being the on-shell
energies of the i-th nucleon after and before the meson exchange,
respectively. As pointed out above, this defines the potential only
on-energy-shell. However, the Lippmann-Schwinger equation and even
the first order Born approximation to the wave function require
the potential off-energy-shell. The extended S-matrix approach
\cite{ATA} defines the most general off-energy-shell continuation
of $V_\sigma$ as a class of unitarily equivalent potentials
parameterized by the ``retardation parameter'' $\nu$. The
particular choice $\nu=1/2$ leads to the static potential in the
$NN$ c.m.\ frame. This choice corresponds to the substitution
$\Delta^2= (\Delta_1-\Delta_2)^2/4$. Most realistic $NN$
potentials, namely those fitted to the data below pion threshold,
in particular the Bonn B potential used in this paper, are
energy-independent and static in the nucleon c.m.\ frame and can
be therefore considered to be consistent with this
construction.

For the pion re-scattering diagram S-matrix prescription leads to
a single effective operator (both for FSI and ISI diagrams) of the
form:
\begin{equation}
 \hat{O}_{rs}^{S}=
 \frac{f(\Omega)}{m_\pi^2+ \vec{q}^{\, \prime 2}-(\Omega)^2} \, ,
\label{msmat}
\end{equation}
where $\Omega= \epsilon'_2- \epsilon_2= \epsilon_1- \epsilon'_1$.

Let us finally introduce approximations to the time-ordered PT result given
by Eqs.~(\ref{afsi}-\ref{vsig}) and explain how the S-matrix technique fits
in. The first of them is {\em the on-shell approximation}, in which the two
nucleons in the intermediate state are put on-energy shell. That is, we put
$\omega_1+ \omega_2= F_1+ F_2$, which also implies $E_1+E_2=\omega_1+
\omega_2+ E_{\pi}$. For the scalar potential in the FSI diagram this leads to
(\ref{vsigos}), where now $\Delta= F_1-\omega_1=\omega_2-F_2$ is the energy
transfer in the corresponding vertices.

For the re-scattering operator (\ref{mfsi}) the on-shell replacement implies
\begin{equation}
 \hat{O}_{rs} \rightarrow \hat{O}_{rs}^{on} =
-\frac{1}{2\omega_\pi}
 \left[ \frac{f(\omega_\pi)}{E_2-\omega_2-\omega_\pi}+
 \frac{f(-\omega_\pi)}{\omega_2-E_2-\omega_\pi} \right] \, .
\label{mfsios}
\end{equation}
Clearly, the S-matrix pion re-scattering operator $\hat{O}_{rs}^{S}$ follows
from  $\hat{O}_{rs}^{os}$  if one assumes the energy conservation at each
vertex. The exchanged pion is then no longer on-mass-shell and we have to
replace $f(\omega_\pi) \rightarrow f(E_2-\omega_2)$ and $f(-\omega_\pi)
\rightarrow f(E_2-\omega_2)$: in the first time-ordered diagram the virtual
pion is entering the re-scattering vertex and in the second one it is emitted
from this vertex (as defined on Fig.~\ref{diagrams2}).

The on-shell approximation as introduced in Refs.~\cite{H,S} actually
coincides with the S-matrix approximation defined above. The re-scattering
operator (\ref{msmat}) can be obtained directly from (\ref{vtil}) and
(\ref{gpi}) by the substitutions following from the on-energy-shell prescription
and the energy conservation in individual vertices
$\omega_\pi=E_2-\omega_2=-\left(E_1-\omega_1-E_\pi \right)$ as follows:
\begin{equation}
 \hat{O}_{rs} \rightarrow -\frac{1}{2 \left(E_2-\omega_2 \right)}
 \frac{-2 \left(E_2-\omega_2 \right) \times  f\left(E_2-\omega_2 \right) +
 0 \times f \left(\omega_2-E_2 \right)}{\left(E_2-\omega_2-\omega_\pi \right)
 \left( \omega_2 -E_2-\omega_\pi \right)}=\hat{O}_{rs}^{S} \, .
\label{vfullon}
\end{equation}

In Ref.~\cite{MP} the extra kinematical factors in (\ref{vtil}) multiplying
the function $f\left(\omega_\pi \right)$ were interpreted as form factors and
kept unaltered, i.e., the substitution above was made only in $G_\pi$ and
$f(\omega_\pi)$ of (\ref{mfsi}), not in the kinematical factors included in
the function $\tilde{f}$.

In equation (\ref{vfullon}) the effective pion propagator  $G_\pi$ is
seen to take its Klein-Gordon form:
\begin{equation}
G_\pi \rightarrow G_\pi^{on}= 1/[(E_2-\omega_2)^2-\omega_\pi^2] \,
. \label{gpion}
\end{equation}
The replacement (\ref{gpion}) does not significantly alter the results, as we
will show below, but the corresponding substitution alone in the pion re-scattering
vertex $f(\pm\omega_\pi) \rightarrow f(\pm(E_2-\omega_2))$ present in
Eq.~(\ref{vtil}) leads to a large enhancement of the amplitude (\ref{afsi}).
According to Ref.~\cite{S} it increases the cross section by almost factor of
3. Since the splitting of $\hat{O}_{rs}$ is not defined unambiguously, in the
work reported here
the on-shell replacement is made in the whole re-scattering operator. We
demonstrate in this paper that making the on-shell replacement in the whole
operator leads to a significantly smaller deviation from the
reference result, compared to the replacement $f(\pm\omega_\pi) \rightarrow
f(\pm(E_2-\omega_2))$ involving $f$ only .

In our previous paper \cite{MP} we considered also other
approximations (besides the {\em on-shell} one): the so-called
{\em static} and {\em fixed threshold-kinematics} approximations,
defined by the replacement of the energy of the exchanged pion
$E_2-\omega_2$ by zero and $m_\pi/2$ (its threshold value),
respectively. The static approximation was in Ref.~\cite{MP}
considered only for $G_\pi$, the fixed threshold-kinematics one
for $G_\pi$ and also for $f(\pm\omega_\pi)$ (the additional
kinematical factors in $\tilde{f}$ were again kept unchanged). In
this paper we again make these approximations in the full
re-scattering operator (\ref{msmat}), replacing $\Omega
\rightarrow 0$ and $\Omega \rightarrow m_\pi/2$, respectively. For
$V_\sigma$ the static  approximation is defined by $\Delta
\rightarrow 0$.

\section{Results}

For numerical calculations we consider the $NN \rightarrow(NN)\pi$ transition
in partial waves  $^3P_0 \rightarrow (^1S_0)s_0$. Amplitudes and cross
sections are evaluated both with the simple interaction $V_\sigma$ and with
the Bonn B potential. We test the S-matrix prescription for the re-scattering
operator (\ref{msmat}) and also the fixed threshold-kinematics and the static
approximations discussed in the previous section. Besides, in order to
compare to the previous papers we include also the results for the on-shell
(\ref{gpion}), fixed threshold-kinematics and the static approximations for
the effective pion propagator $G_\pi$.

\begin{figure}
\includegraphics[width=.98\textwidth,keepaspectratio]{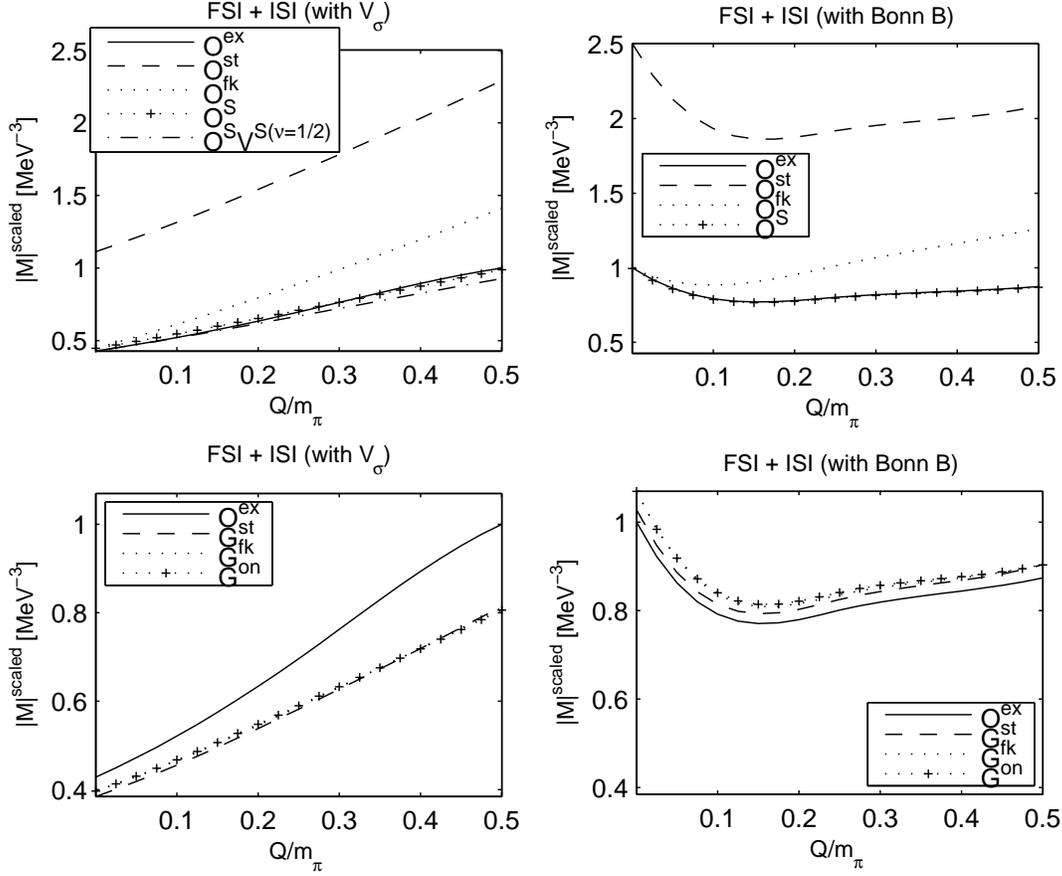}
\caption{Absolute values of the FSI + ISI amplitude as a function
of the excess energy $Q=2E-2M-E_\pi$ (in units of $m_\pi$). The
right(left) panels correspond to the amplitudes with $\sigma$
exchange (Bonn B) for the $NN$ interaction. The amplitudes are
taken at the maximum pion momentum $q_\pi^{max}$, determined by
$Q$. The upper
panels correspond to approximations for the whole operator $
\hat{O}_{rs}$; the lower panels to approximations for the pion
propagator $G_\pi$ only. The solid line $(O^{ex})$ is the reference calculation.
The dashed,
dotted and cross-dotted lines correspond to the static,
fixed threshold-kinematics and on-shell approximations, respectively. The
corresponding operators are $O^{st}$, $O^{fk}$, $O^{S}$  and $G^{st}$, $G^{fk}$,
$G^{on}$. All amplitudes were normalized by a
factor defined by the maximum value of the reference result. }
\label{belowboth}
\end{figure}

In Fig.~\ref{belowboth} we show that the amplitudes with the S-matrix
operator $O^S$ (dotted line with crosses on the upper panels) are the closest
to the reference result (solid line). Using the same approach both for the
operator and for $V_\sigma$ increases slightly the gap from the reference
result (dashed-dotted line versus solid line on the upper left panel). The
fixed threshold-kinematics version of $ \hat{O}_{rs}$, denoted as
$\hat{O}^{fk}$, works well for small values of the excess energy
$Q=2E-2M-E_\pi$, but starts to deviate rapidly with increasing $Q$ (dotted
line in the upper panels of Fig.~\ref{belowboth}). The static approximation
for the re-scattering operator ($\hat{O}^{st}$) overestimates significantly
the amplitude (\ref{afsi}) (dashed versus solid lines in the upper panels).

From the lower panels on Fig.~\ref{belowboth} one sees that all considered
approximations taken only for the effective pion propagator do not differ
much from each other. This had already been found on Ref.~\cite{MP}. It means
that the choices for the energy of the exchanged pion in the effective
propagator alone are not very decisive (solid line versus dotted, dashed and
cross-dotted lines). We notice however that there is a considerable deviation
(dependent on the $NN$ interaction employed)  of all these approximations
from the reference result.

In order to understand this we considered the expansion of the effective pion
propagator $G_\pi$ in Eq.~$\left(\ref{gpi}\right)$ in terms of an
``off-mass-shell" dimensionless parameter $y$:
\begin{equation}
y=-\frac{2 E-E_\pi-\omega_1-\omega_2}{\omega_1-\omega_2+E_\pi}
\, ,
\end{equation}
which measures the deviation of the total energy from the energy  of the
intermediate state with all three particles  on-mass-shell. This Taylor
series expansion gives insight on the small effect of retardation effects in
the propagator, and it reads
\begin{equation}
{G_\pi}=\underbrace{\frac{1}{\left(\frac{E_\pi+\omega_1-\omega_2}{2}\right)^2-
\omega_\pi^2}}_{G^{(1)}_{Tay}}
\left[1+\frac{\left(-2E+E_\pi+\omega_1+\omega_2\right)}
{\left(\frac{E_\pi+\omega_1-\omega_2}{2}\right)^2-\omega_\pi^2}
+...\right] \label{gexpfsi}
\end{equation}
where $G^{(1)}_{Tay}$ has the form of the usual Klein-Gordon propagator.

We notice here that in the case of the ISI amplitude, the representation of
the pion propagator $G_{\pi}$ by its Taylor series, the first term of which
is $G^{(1)}_{Tay}$, fails due to the presence of a pole in the propagator.

\begin{figure}
\includegraphics[width=.70\textwidth,keepaspectratio]{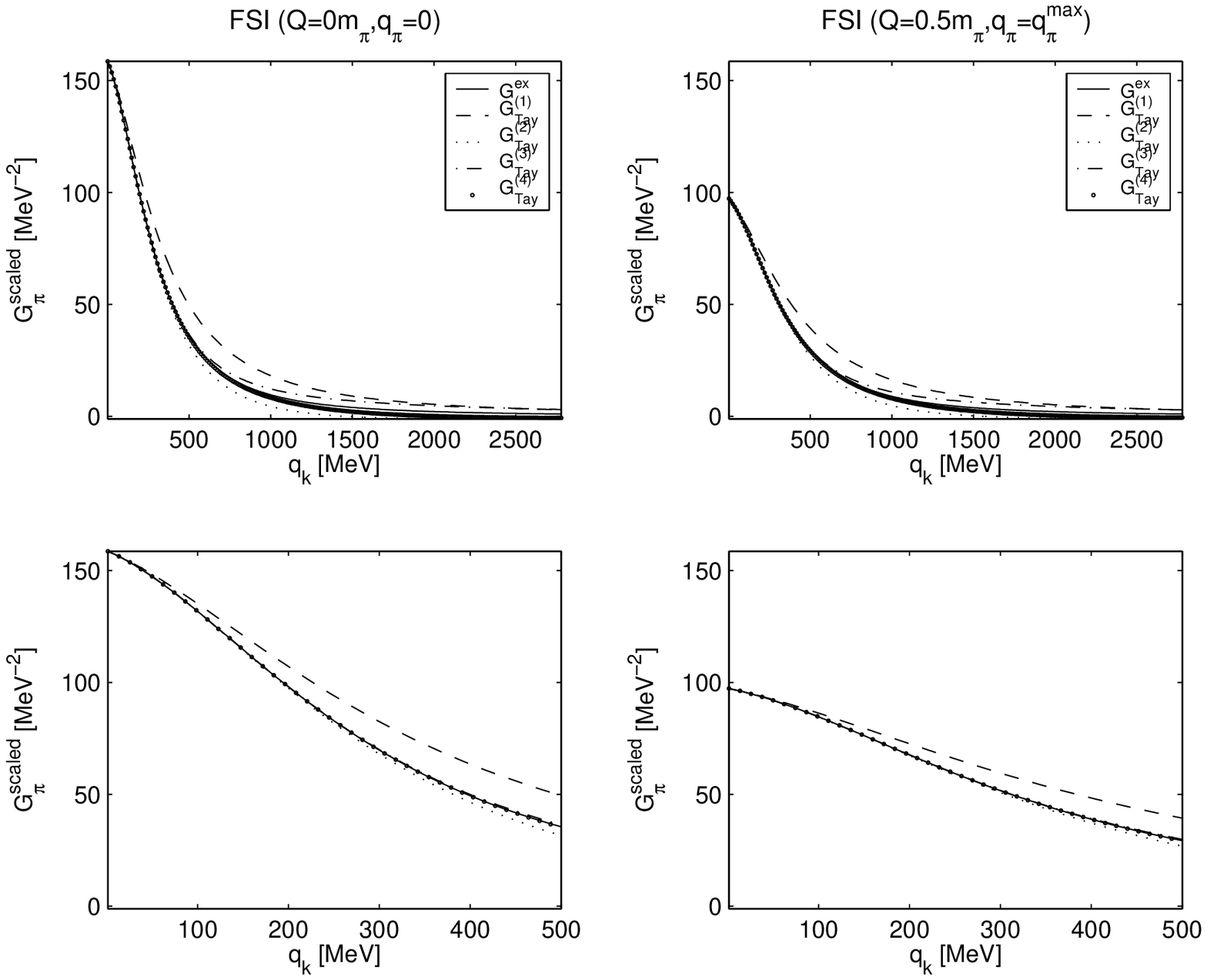}
\caption{Convergence of the Taylor expansion for the pion
propagator $G_{\pi}$ in the FSI amplitude, as a function of the
two nucleon relative momentum (Eq. $\left(\ref{gexpfsi} \right)$).
Left panel: at threshold; right panel: above threshold at maximum
pion momentum for an excess energy $Q$ of $0.5m_\pi$. Bottom panels
zoom into the region of low relative momentum.}
\label{gexpansion}
\includegraphics[width=.66\textwidth,keepaspectratio]{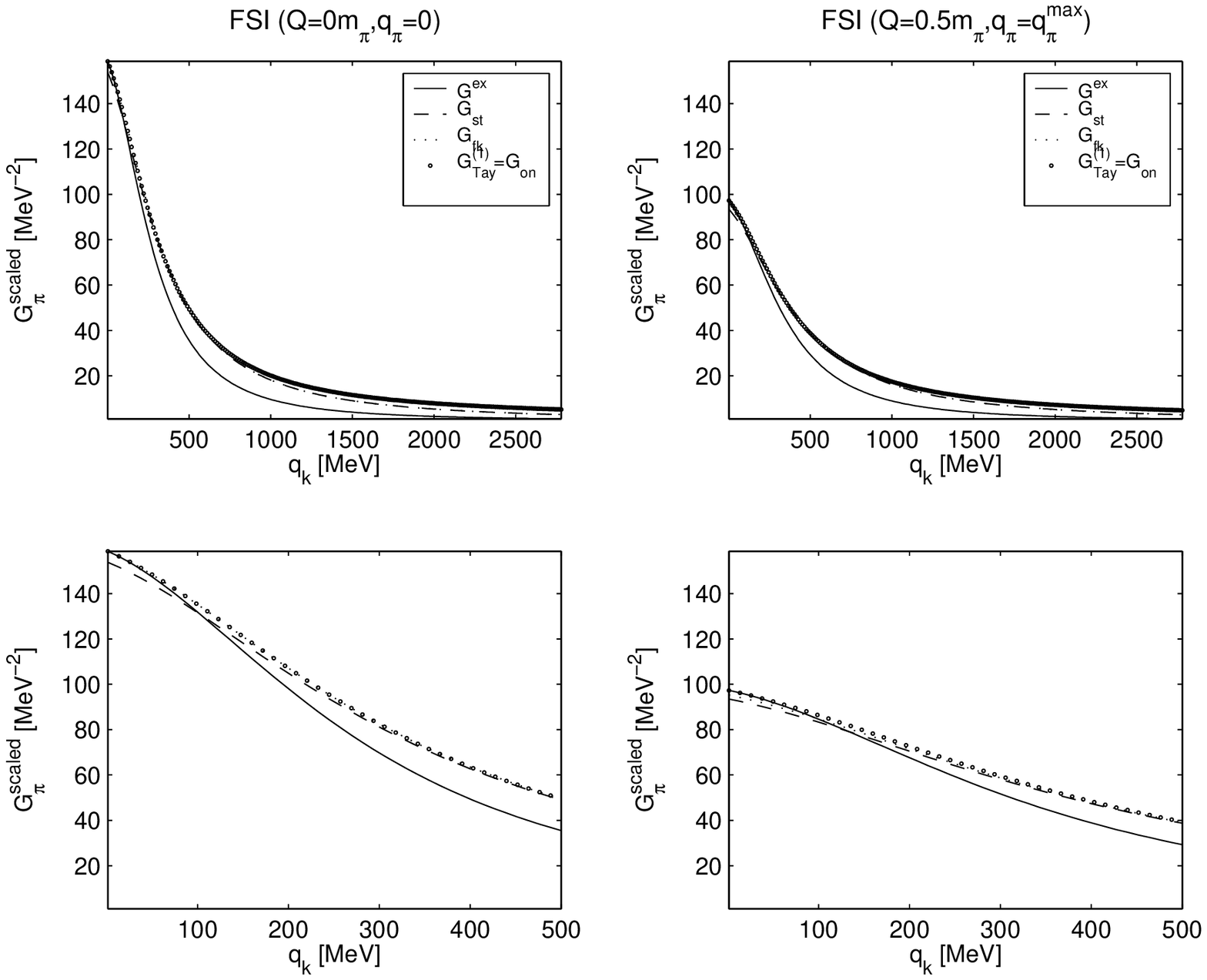}
\caption{Approximations for $G_{\pi}$ in the FSI diagram and the
first term of the Taylor series. Left and right panels with the
same meaning as on Fig.~\ref{gexpansion}.} \label{gapprox}
\end{figure}

Figure~\ref{gexpansion} compares the first four terms $G_{Tay}^{(i)} (i=1,...,4)$
of this expansion with the full effective propagator in Eq.~(\ref{gpi}), as a
function of the two nucleon relative momentum $q_k$, for two different values
of the excess energy $Q=2E-2M-m_\pi$. The convergence of the series demands
at least 4 terms. Besides, as expected, this convergence is momentum-dependent.
We have also compared the first term of this expansion with the already
considered on-shell, fixed threshold-kinematics and static approximations for
the pion propagator. These results are shown on Fig.~\ref{gapprox}. We
realize that all these approximations are very near to the 1st order term of
the Taylor series. The corrections arising from higher order terms in the
expansion are negligible only for low momentum transfer, more precisely in
the range $q_k< 100$MeV.

The deviations of $G^{st}$, $G^{fk}$ and $G^{on}$ from the effective
propagator $G_\pi$ given by  Eq.~(\ref{gpi}) cannot explain the relatively
large deviations obtained on the bottom-left panel of Fig.~\ref{belowboth}
between considered approximations and the reference result. These deviations
follow from the ISI contribution. The weight of the ISI term depends on the
$NN$ interaction employed. It is comparable to the FSI term for $V_\sigma$ (for
which the deviations are large, as seen on the bottom-left panel of
Fig.~\ref{belowboth}), but it is much less important for the full Bonn B
potential (and therefore the corresponding deviations on the bottom-right panel of
Fig.~\ref{belowboth} are indeed much smaller).

\begin{figure}
\includegraphics[width=.98\textwidth,keepaspectratio]{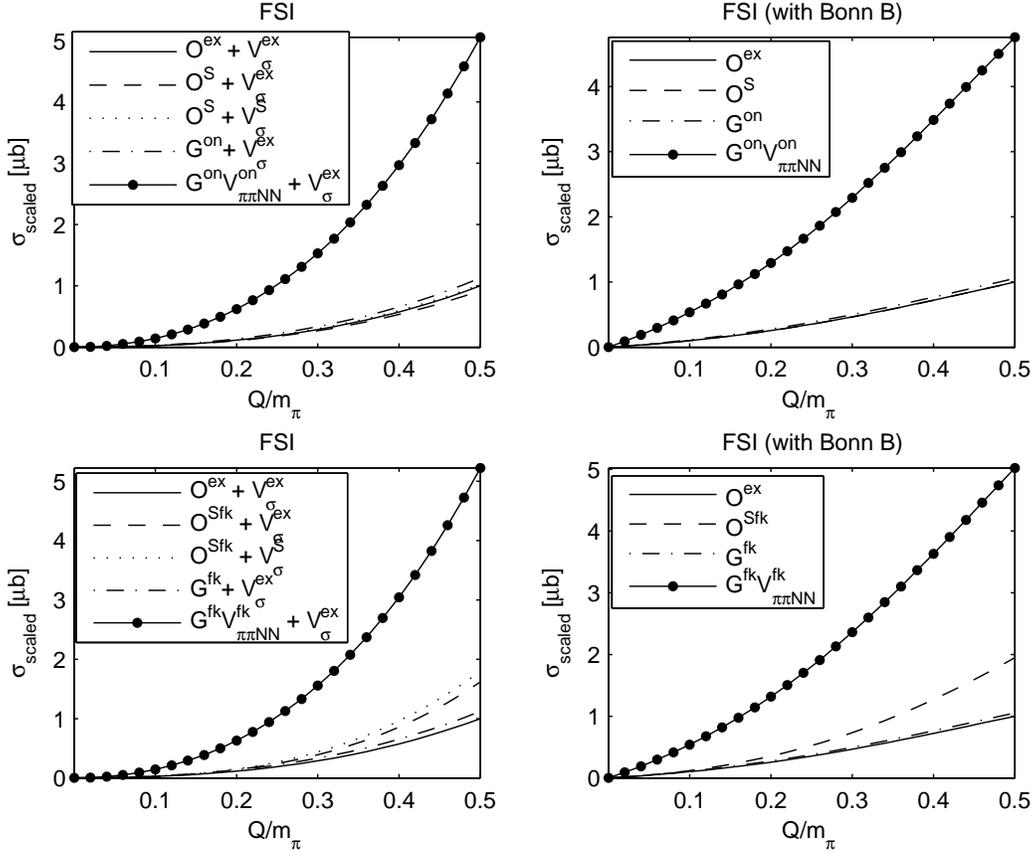}
\caption{Effects of the approximations for the re-scattering
operator $\hat{O}_{rs}$ and for the effective pion propagator
$G_\pi$ as a function of the excess energy $Q$. The cross section
curves shown correspond to the FSI amplitude alone. The upper
panels correspond to the re-scattering operator $\hat{O}$ given by
Eq. $\left(\ref{msmat} \right)$ and the lower panels to fixed
threshold-kinematics approximation. The solid line is the
reference calculation $\left(\ref{afsi} \right)$. The dashed line
is the S-matrix calculation for the re-scattering operator
$\hat{O}$ given by Eq. $\left(\ref{msmat} \right)$ (upper panels)
and the fixed threshold-kinematics approximation for
$\left(\ref{msmat}\right)$ (lower panels). The dotted line
corresponds to take the S-matrix approximation not only for
$\hat{O}$, but also for the $\sigma$-exchange interaction
(\ref{vsigos}). The dashed-doted line corresponds to the on-shell
(upper panels) and  fixed threshold-kinematics (lower panels) prescriptions
only for $G_\pi$. The solid lines with bullets refer to the energy
prescriptions taken for the propagator $G_\pi$ and for
$f(\omega_\pi)$ in Eq.~(\ref{vtil}), as
in \cite{MP}, but not for extra kinematical factors in
$\tilde{f}$. All the cross sections were normalized with a factor
defined by the maximum value of the reference result.} \label{fsi}
\end{figure}

All the findings for the amplitudes manifest themselves also in the results
for the cross section. We show in Fig.~\ref{fsi} the effects of the
considered approximations on the cross section, first taking only the FSI
contribution. On the left panel the amplitude includes $V_\sigma$ for the
$NN$ interaction, on the right panel the Bonn B potential is used. The curves
compare the reference result (solid line in all panels) with the S-matrix
results (upper panels) and  their fixed threshold-kinematics version (lower
panels). The S-matrix approach (dashed  line) is the closest to the reference
result (upper panels of Fig.~\ref{fsi}).

For the case of the $NN$ interaction described by $V_\sigma$ we also show the
result following from the S-matrix prescription applied to the $NN$
interaction (dotted line on left panels in Fig.~\ref{fsi}). For the fixed
threshold-kinematics versions (bottom-left panel) the deviations from the
reference result increase more pronouncedly with the excess energy $Q$, as
expected. The approximations for the energy of the exchanged pion taken in
the pion propagator $G_\pi$ and in the re-scattering vertex $f(\omega_\pi)$,
but not in the kinematical factors of $\left(\ref{vtil} \right)$,
overestimate the cross section by a factor of 5 (solid line with bullets).

\begin{figure}
\includegraphics[width=.98\textwidth,keepaspectratio]{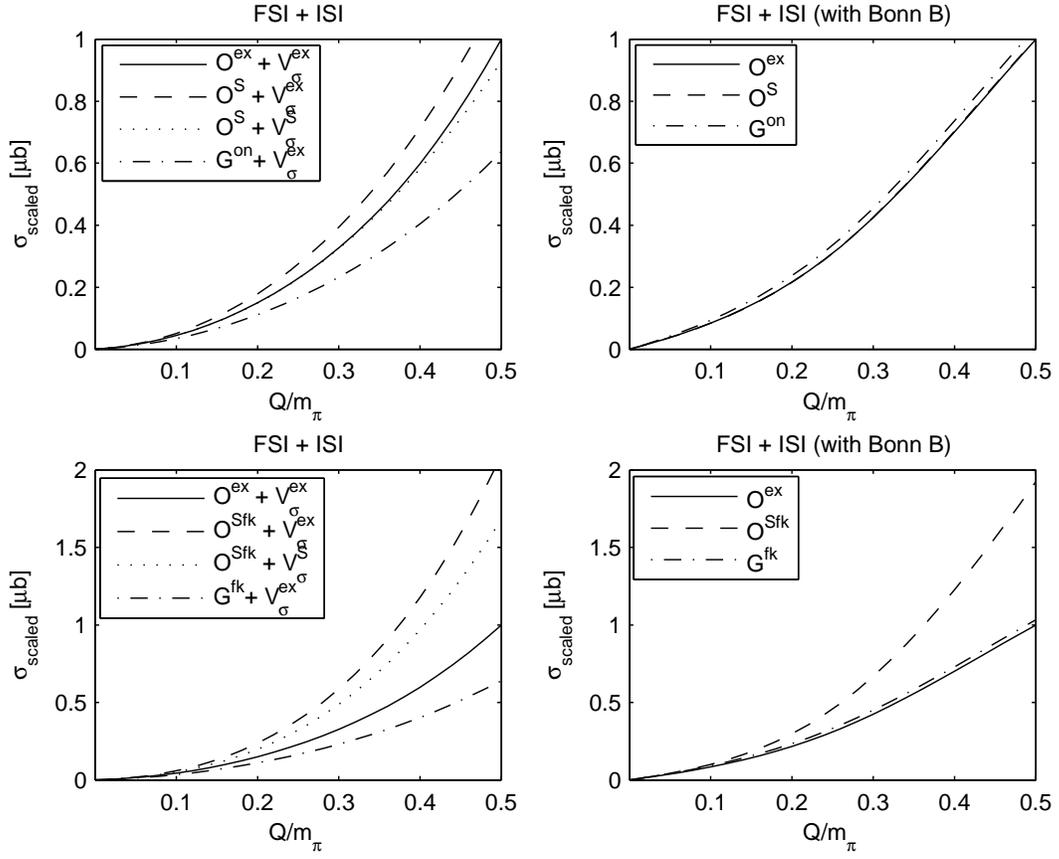}
\caption{The same of Fig. \ref{fsi},  but for the total (FSI+ISI)
cross-section and considering only the
approximations for $\hat{O}$ and $G_\pi$.} \label{bothzoom}
\end{figure}

Finally, we present on Fig.~\ref{bothzoom} the comparison between the
approximated total cross sections with both FSI and ISI included. The
approximation dictated by the S-matrix approach (dashed and dotted lines on
the upper panels of Fig.~\ref{bothzoom}) is clearly seen as the best one.
For the Bonn potential calculation, it practically coincides with the
reference result. As shown in the previous section, this procedure amounts to
extend the on-shell approximation, used in Ref.~\cite{MP} for $G_\pi$ and
$f(\omega_\pi)$ alone, also to the multiplicative kinematical
factors showing up in the operator $\tilde{f}$ (see
Eqs.~(\ref{mfsi}-\ref{gpi})).

To conclude we notice, moreover, that for the
realistic $NN$ interaction, the difference between the S-matrix
approach (upper right panel on Fig.~\ref{bothzoom}) and its
fixed threshold-kinematics version (lower right panel of the same figure) is
not very important near threshold, provided the excess energy does
not exceed $\approx 30$ MeV ($Q/m_\pi \sim 0.2$).

\section{Conclusions}

1) The usual approximations to the effective pion propagator \cite{MP,S,PCK},
obtained from a quantum-mechanical reduction of the Feynman diagram
describing the re-scattering, which are rather close to the first order
term of a Taylor series in a parameter measuring off-mass-shell effects in
the intermediate states. The series converges rapidly  for the FSI amplitude
near threshold. As a consequence, retardation effects are not decisive in the
pion re-scattering mechanism near the threshold energy for pion production.

2) As for the pion energy in the $\pi N$  re-scattering amplitude, the
on-shell approach when used only in $f(\omega_\pi)$  overestimates
significantly  the reference result. Nevertheless, and this is the key point
of this paper, this deviation is dramatically reduced if the approximation
coming from the S-matrix approach is used consistently in the whole effective
operator. This procedure amounts to extend the on-shell approximation used in
Ref.~\cite{MP} for $G_\pi$ and $f (\omega_\pi)$ to the full
operator $\tilde{f}$, including kinematical factors which differently weight
the two dominant time-ordered diagrams. The amplitudes and cross sections
obtained with the S-matrix effective operator are very close to those
obtained with the time-ordered one in the considered kinematical region.

The re-scattering operator of this paper for the neutral pion
production in the isoscalar $\pi$N channel thus indeed seems to be relatively
unimportant: its enhancement reported in previous papers followed from
inconsistent or too crude (static or fixed threshold-kinematics) treatment of
the energy dependence of the effective operator.  Our findings explain why
the calculation of Ref.~\cite{S}, where the on-shell approximation is used,
artificially enhances the contribution of the isoscalar re-scattering term.
On the other hand, importantly and in retrospect, our results support the
choice done in Refs.~\cite{PRS,ASP,PCK} for the different production
operators considered.

The re-scattering mechanism is filtered differently by other
spin/isospin channels in pion production reactions. For charged
pion production reactions the general irreducible re-scattering
operator comprises also the dominant isovector Weinberg-Tomozawa
term of the $\pi$N amplitude, and its importance is therefore
enhanced. Investigation of these channels within the approach
of this paper is in progress.

\begin{acknowledgments}
J.A. was supported by the grant  GA CR 202/03/0210 and by the
projects K1048102, ASCR AV0Z1048901.
 M.T.P. was supported by the grant CERN/FIS/43709/2001 and
V.M. was supported by FCT under the grant SFRD/BD/4876/2001.
\end{acknowledgments}

\newpage

\end{document}